\documentclass[aps,prb,preprint,groupedaddress]{revtex4-2}

\usepackage{graphicx}
\usepackage{amsmath,amsthm,amssymb,amsfonts}

\begin{document}


\title{First-principle calculations of plasmon excitations in graphene, silicene and germanene}


\author{Pengfei ~Li}
\email[]{lpf91124@xaut.edu.cn}
\affiliation{School of Science, Xi’an University of Technology, Xi’an, 710048, China}

\author{Rong ~Shi}
\affiliation{CAS Key Laboratory of Quantum Information, University of Science and Technology of China, Hefei 230026, Anhui, China}

\author{Peize ~Lin}
\affiliation{Institute of Physics, Chinese Academy of Sciences, Beijing 100190, China}
\affiliation{Songshan Lake Materials Laboratory, Dongguan 523808, Guangdong, China}

\author{Xinguo ~Ren}
\email[]{renxg@iphy.ac.cn}
\affiliation{Institute of Physics, Chinese Academy of Sciences, Beijing 100190, China}
\affiliation{Songshan Lake Materials Laboratory, Dongguan 523808, Guangdong, China}


\date{\today}

\begin{abstract}
Plasmon excitations in graphene, silicene and germanene are studied using linear-response time-dependent density functional theory within the random phase approximation (RPA). In this work, we examine both the plasmon dispersion behavior and lifetime of extrinsic and intrinsic plasmons for these three materials. For extrinsic plasmons, we found that their properties are closely related to Landau damping. In the region without single-particle excitation (SPE), the plasmon dispersion shows a $\sqrt{q}$ behavior and the lifetime is infinite at the RPA level, while in the single-particle excitation region, the plasmon dispersion shows a quasilinear behavior and the lifetime is finite. Moreover, for intrinsic plasmons, unlike graphene, the plasmon dispersion behavior of silicene and germanene exhibits a two-peak structure, which can be attributed to the complex and hybridized band structure of these two materials.
\end{abstract}


\maketitle



\section{\label{sec:intro}Introduction}
In recent years, two-dimensional (2D) hexagonal materials arranged in a honeycomb lattice has been the subject of intensive edge-cutting researches in condensed matter physics, materials science and engineering \cite{novoselov2004electric,geim2009graphene,neto2009electronic}. Graphene is the most prominent example
of this type of materials, however, the incompatibility of carbon-based materials with current silicon-based or germanium-based electronics makes it currently unsuitable for practical use. Therefore, considerable current research interests have been extended to other candidates    
among the group-IV elements, like 
silicene and germanene -- the counterparts of graphene for Si and Ge elements \cite{houssa2011electronic,vogt2012silicene,o2012stable,cai2013stability,kaloni2016current,molle2018silicene,guo2020two,zhang20182d,ma2021two,chia2021functionalized,ang2021effect,zheng2021multichannel}. 
Although C, Si and Ge belong to the same group (Group IV) in the periodic table, their chemical properties, largely governed by orbital hybridizations, are substantially different\cite{houssa2011electronic}. The energy required for hybridization of $s$ and $p$ orbitals in C is much larger than that in Si and Ge, and hence the former prefers $sp^2$ hybridization more than the latter two. As a consequence, all the carbon atoms in graphene prefer to stay in the same plane, whereas the Si and Ge atoms in silicene and germanene prefers mixed $sp^2$-$sp^3$ hybridization, so that the two atoms in primitive cell are not any more in the same plane. In anothor word, silicene and germanene adopt buckled honeycomb lattice structure.

The strong coupling between the plasmon modes and light or charged particles has received a lot of attentions. 
With small spatial extension and huge optical enhancements, these 
plasmonics can be used to design terahertz metamaterials\cite{ju2011graphene}, modulators\cite{sensale2012broadband},
field detectors\cite{salamin2019compact} and biosensings\cite{ahmadivand2020terahertz}.
Recent years, many studies have been focused on the plasmon dispersions of monolayer graphene\cite{grigorenko2012graphene,pisarra2014acoustic,liou2015pi,mencarelli2015spatial,li2017first}, bilayer graphene\cite{fei2015tunneling,pisarra2016dielectric} and graphene nanoribbons\cite{fei2015edge,gomez2016plasmon,gomez2017tunable}. Besides these, several studies are focused on the monolayer\cite{mohan2014electronic,gomez2017plasmon,sindona2018interband} and bilayer\cite{mohan2013first} silicene, but
the plasmon properties in germanene are so far only considered within tight-banding models \cite{shih2017coulomb,shih2018coulomb,iurov2017temperature}, the results under first-principles framework, have not been 
reported so far. Derived from their unique electronic band structures, the plasmon excitation spectra of these three materials are very similar. At low energies under finite electron/hole dopings, the so-called Dirac plasmon appears, originating from the intraband transitions in the vicinity of K points of the Brillouin zone (BZ)\cite{wunsch2006dynamical,hwang2007dielectric}. Because the formation of Dirac plasmons requires intervention from external means, like chemical dopings or by electric gating, therefore, we call it extrinsic plasmon. At higher energy regimes, there exist intrinsic $\pi$-like plasmons. In graphene, they mainly arise from the collective excitations of electrons from $\pi$ to $\pi^\ast$ bands around the M points\cite{marinopoulos2004ab}, while in silicene and germanene, due to the band structure hybridization, a mixed transition from $\pi$ to $\pi^\ast$ or $\sigma^\ast$ predominant hybridizaiton bands form the $\pi$-like plasmons\cite{gomez2017plasmon}. At even higher energies, the bands far above the Fermi level, which are mainly composed of $\sigma$-type bands start to contribute, and the excitations formed by the transition between $\pi$ and $\sigma$ predominant bands are called $\pi$-$\sigma$ plamsons. Corresponding to the extrinsic plasmons,  we call the latter two plasmons intrinsic plasmon.  Especially, for Dirac plasmons, because of its strong field localization, low energy loss, and high tunability in frequency range by varying the doping level, they are expected to play a key role in the design of next-generation nanophotonic and nanoelectronic devices \cite{jablan2009plasmonics,low2014graphene}. In order to design high-performance devices, an in-depth understanding of the plasmon properties of graphene, silicene and germanene is indispensable. In this work, we presented a detailed 
first-principles study of the plasmon properties of graphene, silicene and germanene, including their intensities, dispersion relationships and lifetimes.
Particular attention is paid to the influence brought by the doping level (charge carrier concentration).
A systematic comparison of the plasmon properties across these three systems is particularly inspiring, as the underlying relationships between
the element type, geometrical and electronic structures, and the plasmon properties can be revealed from these studies.

In this paper, we use the time-dependent density-functional theory (TDDFT) within the random-phase approximation (RPA) framework \cite{yuan2009linear} to study the plasmon properties of graphene, silicene and germanene. Compared to the earlier tight-binding model analysis \cite{wunsch2006dynamical,hwang2007dielectric,jablan2009plasmonics}, which is valid only in the vicinity of the K points in the BZ, the \textit{ab initio} TDDFT method can offer a unified description of all types of plasmon excitations, including the extrinsic and intrinsic plasmons. In this work, we also discuss some computational details. The first issue is how to describe the form of the dielectric function for 2D systems. Here we showed two schemes, their difference lies in the form of the Coulomb potential. Although both methods give the same results on plasmon excitation energies and lifetimes, their real parts for the inverse of the dielectric functions are different, may cause an impact on judging whether it is plasmon or not. Moreover, we also analysis the influence of choosing different broadening parameter $\eta$ in the non-interacting response function $\chi_0$ calculation.

The rest of the paper is organized as follows. In Sec.~\ref{sec:methods} the basic equations of the TDDFT-RPA approach are presented. In Sec.~\ref{sec:details} the main parameters for ground states and TDDFT-RPA calculations are given. The calculated results for extrinsic and intrinsic plasmons for graphene, silicene and germanene are presented in Sec.~\ref{sec:results}. Finally we draw a conclusion and present our perspectives in Sec.~\ref{sec:conclusion}.

\section{\label{sec:methods}Methods}
In this section, we briefly recapitulate the key equations behind the TDDFT-RPA approach adopted in the present work\cite{li2017first}.
The basic step is to construct the non-interacting response function $\chi^0$, which describes the density response of
the Kohn-Sham system in the linear regime. Here, we consider introducing an electron or a photon, with incident momentum ${\bf q}$ and frequency $\omega$, which weakly perturbs the KS system. The non-interacting response function $\chi^0$ is given by the well-known Adler-Wiser formula\cite{Adler:1962,Wiser:1963}, which in the reciprocal space is expressed as follows,
\begin{widetext}
\begin{equation}
\begin{aligned}
\chi^0_{\bf{G},\bf{G}^\prime} ({\bf q}, \omega) = \frac{1}{\Omega} \sum_{\bf k}^{1BZ} \sum_{n,n^\prime = c,v} 
&\frac {f_{n,\bf{k}}-f_{n^\prime,\bf{k}+\bf{q}}} {{\omega + \epsilon_{n,\bf{k}} -\epsilon_{n^\prime,\bf{k}+\bf{q}} + i\eta}}  \\
&\times \langle n,\bf{k}|e^{-i(\bf{q}+\bf{G})\bf{r}}|n^\prime,\bf{k}+\bf{q}\rangle \langle n^\prime,\bf{k}+\bf{q}|e^{i(\bf{q}+\bf{G}^\prime)\bf{r}^\prime}|n,\bf{k}\rangle\, . 
\end{aligned}
\label{eq:chi_0}
\end{equation}
\end{widetext}
In Eq.~(\ref{eq:chi_0}) , $\Omega$ stands for the volume of the cell, while $f_{n,k}$, $\epsilon_{n,k}$, $|n,\bf{k} \rangle$
are the Fermi occupation numbers, KS eigenvalues \cite{Kohn/Sham:1965} and eigenvectors, respectively.
The summation over ${\bf k}$ goes over the first Brillouin zone (1BZ), and $\eta$ represents the broadening parameter. The influence of different 
choices of $\eta$ on the obtained results has not been thoroughly discussed.  In Appendix.~\ref{sec:eta}, this issue will be analyzed in depth regarding
the plasmon peak positions as well as lifetimes when different values of $\eta$ are chosen.
The details of the implementation within a plane-wave basis set is given in Ref.~\onlinecite{li2017first}.

The next step is to construct the the system’s interacting response function $\chi$. Within the TDDFT framework,  it is linked to the non-interacting counterpart $\chi^0$ via the Dyson equation\cite{1996Excitation}
\begin{equation}
\chi_{\bf{G},\bf{G}^\prime}= \chi_{\bf{G},\bf{G}^\prime}^0 +
(\chi^0 v \chi)_{\bf{G},\bf{G}^\prime},
\label{eq:LR-TDDFT}
\end{equation}
In the RPA, the kernel $v_{\bf{G},\bf{G}^\prime}$ in Eq~(\ref{eq:LR-TDDFT}) is reduced to the static bare Coulomb kernel. For 2D materials, 
we use the periodic boundary conditions in DFT calculations, and the interaction between the periodic replicas (along z direction) is non-negligible. 
In order to cancel out this unphysical interaction arising
from the long-range character of the Coulomb potential, we replace the bare Coulomb kernel by the truncated Coulomb potential\cite{Rozzi/etal:2006}
\begin{equation}
v_{\bf{G},\bf{G}^\prime}(\bar{\bf{q}}) = \frac{4\pi\delta_{\bf{G},\bf{G}^\prime}}{|\bar{\bf{q}}+\bf{G}|^2}
  \left[1 - (-1)^{n_z}e^{-|\bar{\bf{q}}+\bar{\bf{G}}|{\frac{L_z}{2}}}\right]
\label{eq:truncated_Comlomb kernel_simple}
\end{equation}
where $\bf{G} =(\bar{\bf{G}}_1,G_{1,z})$ and $\bf{q}=(\bar{\bf{q}},0)$, with $\bar{\bf{G}}$ and $\bar{\bf{q}}$ being respectively the two-dimensional reciprocal lattice vector and Bloch momentum vector in the basal plane. $L_z$ is the length of the lattice vector in the $z$ direction, and  $n_z=G_z L_z / {2\pi}$  being an integer number.

Finally, the plasmon excitations can be obtained from the imaginary part for the inverse of the dielectric function Im\{$\epsilon^{-1}$\}, without considering the nonlocal field effects,  we can get $\epsilon^{-1}$ by

\begin{equation}
(\epsilon^{-1})_{\bf{G} \bf{G}^\prime} = \delta_{\bf{G} \bf{G}^\prime} + (v\chi)_{\bf{G} \bf{G}^\prime}
\label{eq:inverse_epsilon}
\end{equation}
where $v$ represents the Coulomb potential. In the literature
there exists two different schemes for choosing the form of Coulomb potential in Eq.~\ref{eq:inverse_epsilon}, i.e.,
the 2D-form\cite{nazarov2015electronic} 
\begin{equation}
v = \frac{2\pi L_z}{\bar{q}}
\label{eq:2D_coulomb_kernel_1}
\end{equation}
and the gradual-form\cite{li2017first,gomez2017plasmon}(nearly 2D-form at small q’s and 3D-form at large q’s)
\begin{equation}
v = \frac {4\pi(1-e^{-\bar{q}L_z/2})} {{{\bar{q}}^2}}\, .
\label{eq:2D_coulomb_kernel_2}
\end{equation}
In practical calculations, it is noted that, although both of the two forms give the same results on plasmon excitation energies (the peak positions of Im\{$\epsilon^{-1}$\}) and lifetimes(the full width at half maxima of the plasmon peak of Im\{$\epsilon^{-1}$\}, FWHM), the real part for the inverse
of the dielectric function, Re\{$\epsilon^{-1}$\} shows different situations. In this paper, we use the pure 2D-form for production calculations, and the reason for this choice is elaborated in Appendix.~\ref{sec:discribe_2D}.

In the end, the plasmon structure is provided by the energy-loss function, being proportional to the imaginary part of the inverse permittivity
\begin{equation}
   E_{Loss}=-\text{Im}\{\epsilon^{-1}_{\bf{G}=0,\bf{G}^\prime=0}\} = - \frac {2\pi L_z} {\bar{q}} \text{Im}\{{\chi}_{\bf{G}=0,\bf{G}^\prime=0}\} \, 
  \label{eq:Imchi-1}
\end{equation}

\section{\label{sec:details}Computational Details}
We carried out the DFT calculations using the Atomic-orbital Based Ab-initio Computation at UStc (ABACUS) package\cite{chen2010,li2016large} whereby the norm-conserving pseudopotentials\cite{hamann2013optimized,schlipf2015optimization} is used to describe the ion-electron interactions. The generalized gradient approximation  (GGA) in the form of the Perdew, Burke, and Ernzerhof  (PBE)\cite{perdew1996generalized} was used for the exchange-correlation functional and the Kohn-Sham (KS) electron wave functions are expanded in terms of the plane-wave (PW) basis. After convergence tests, the kinetic energy cutoff of plane wave basis was set to 50 Ry, 20 Ry and 40 Ry for graphene, silicene and germanene, respectively. We utilized the periodic boundary conditions and the Monkhorst-Pack method\cite{monkhorst1976special} with 50$\times$50$\times$1 k-point meshes for these three 2D systems in full-cell structural relaxations. The electronic iteration convergence threshold was set to $10^{-9}$ eV, while structural relaxations were performed until forces on each atom were below 0.01 eV/\AA. During the cell relaxation, the threshold for stress was set to 0.05 GPa. The vacuum separation between periodic images was chosen to be 20 \AA.

In TDDFT-RPA calculations, the Brillouin zone is sampled with 200$\times$200$\times$1, 128$\times$128$\times$1 and 128$\times$128$\times$1 Monkhorst-Pack kpoint grid for graphene, silicene and germanene, respectively. The broadening parameter $\eta$ is set to 0.002 Ry for major plasmon dispersion calculations and the $\chi$ matrix is expanded in terms of 50 ${\bf G}$ 
vectors. For extrinsic plasmon calculations, 12, 12 and 22 bands are used for graphene, silicene and germanene, respectively 
(the germanium pseudopotential contains inner 3d electrons), while for intrinsic plasmon calculations, 20, 50 and 60 bands are used for graphene, silicene and germanene, respectively. 

\section{\label{sec:results}Results and Discussions}
In this section, the computed results of the ground state properties and full plasmon spectra of graphene, silicene and germanene will be presented.
We will discuss both the dispersion relations and lifetimes for the extrinsic plasmons, while for the intrinsic plasmons, only the dispersion behaviors will be involved.

\subsection{DFT calculation}
\begin{figure}[ht]
\centering
\includegraphics[width=1\textwidth]{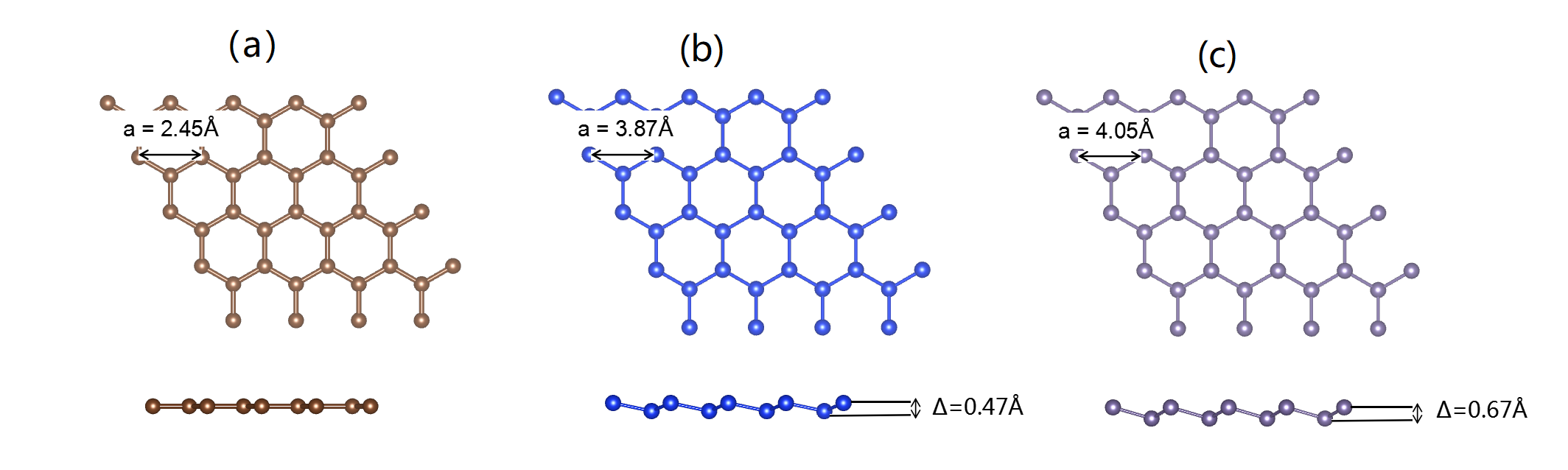}
\caption {The structure and the geometrical parameters of (a) graphene (a = 2.45 \AA), (b) silicene (a = 3.87 \AA, $\Delta$ = 0.47 \AA ) and (c) germanene (a = 4.05 \AA, $\Delta$ = 0.67 \AA).} 
\label{fig:structure}
\end{figure}

After the geometry optimization, the lattice constant $a$ was found to be 2.45 \AA, 3.87 \AA \ and 4.05 \AA \ for graphene, silicene and germanene, respectively. The optimized structures are shown in Fig.~\ref{fig:structure}. One clearly observes that the lattice constant becomes larger as the atomic number increases. In addition, it is also observed that
the two atoms in the graphene primitive cell are in the same plane, while they are not in silicene and germanene. Namely, 
the 2D planes are buckled for silicene and germanene, with the buckling parameters $\Delta$ 
being 0.47 \AA \  and 0.67 \AA, respectively, which is consistent with previous studies reported in the literature \cite{2014Silicene}.

\begin{figure}[ht]
\centering
\includegraphics[width=1\textwidth]{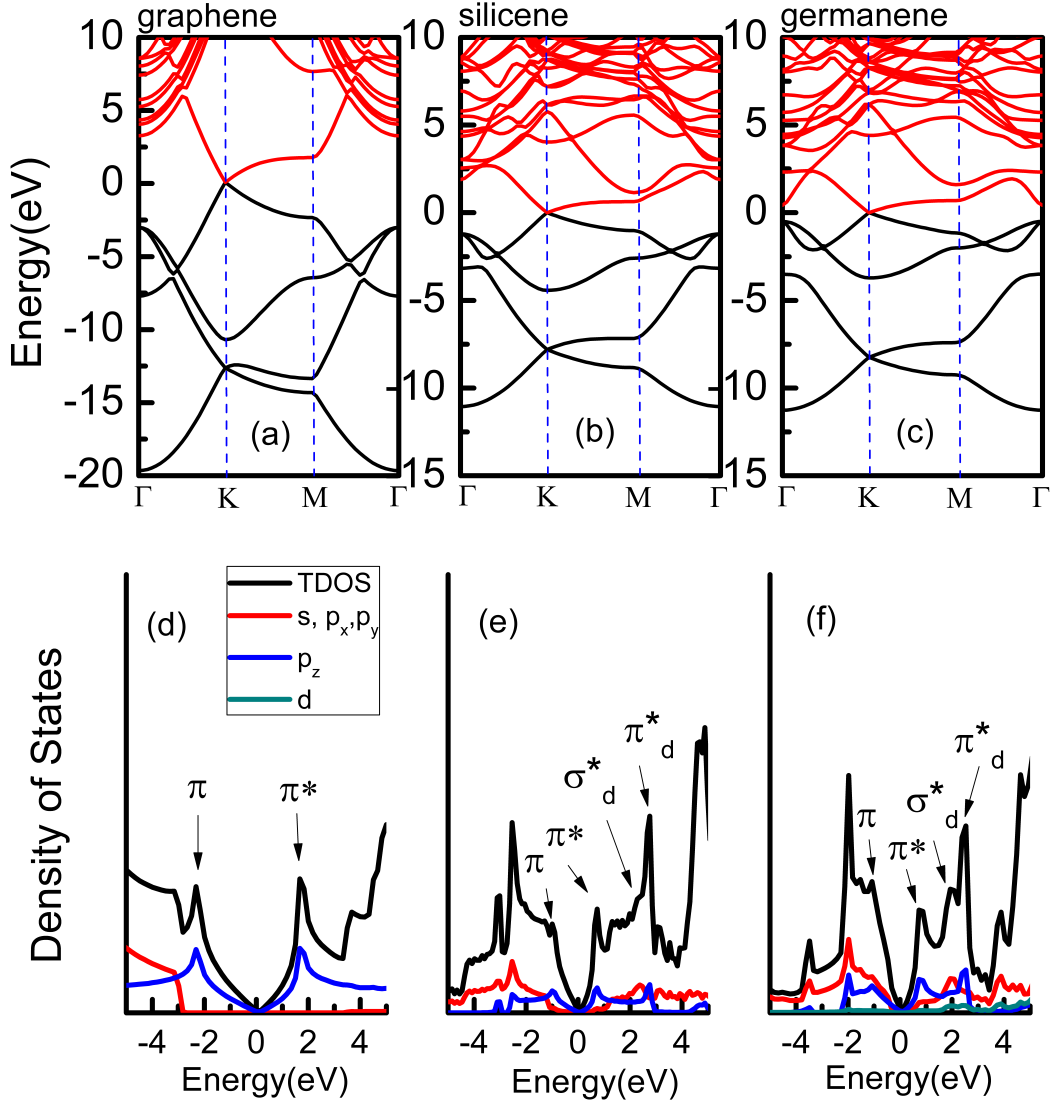}
\caption { The upper panels (a)(b)(c) shows the electronic band structure for graphene, silicene and germanene. The black lines represent the occupied bands and the red lines represent the unoccupied bands. The lower panels (d)(e)(f) shows the total density of states (TDOS) and orbital projected density of states (PDOS) for graphene, silicene and germanene, respectively.} 
\label{fig:band_dos}
\end{figure}

The band structures of graphene, silicene and germanene are plotted in upper panels of Fig.~\ref{fig:band_dos}. We witness all these three hexagonal honeycomblike materials own a Dirac cone structure at the K points near the Fermi level, no matter buckled or not. Compared to graphene, the conduction bands of silicene and germanene are more complicated, coming from stronger hybridization effects. To better analyze the electronic structure of these materials, the total density of states (TDOS) and the orbital projected density of states (PDOS) are represented in the lower panels of Fig.~\ref{fig:band_dos}. As is well known, the $s$, $p_x$, $p_y$ states form the $\sigma$ bands, and the $p_z$ states form the $\pi$ bands. By inspection of Fig.~\ref{fig:band_dos}(d)(e)(f), we can clearly see that, within the energy window from -3 eV to 3 eV, for graphene, all bands are $\pi$ bands. On the contrary, for silicene and germanene, they are hybrid $\pi$ and $\sigma$ bands. In the low-energy region (0-1 eV), where $\sigma$ bands give little contributions, same as graphene, we call it $\pi^\ast$ bands. While in higher energy region (1-3 eV), as the $\sigma$ bands start to contribute, we define another two band types, $\sigma$-predominant bands ($\sigma^\ast_d$) and $\pi$-predominant bands ($\pi^\ast_d$). Compared to silicene, it is noted that the $\sigma^\ast_d$ peak in germanene is so pronounced that it forms an obvious peak in the TDOS curve (Fig.~\ref{fig:band_dos}(f)), which will strongly influences the behavior of the energy loss spectrum, as discussed later in Sec.~\ref{sec:results} C. 

Due to the linear relation between energy and momentum near K point, one can also obtain the Fermi vectors $k_F$ by $k_F = E_F/\hbar v_F$\cite{liu2008}, where $v_F$ represents the Fermi velocity which can be derived from the band structure by $v_F = \frac{1}{\hbar}\frac{\partial E}{\partial k}$.

\subsection{Extrinsic plasmons}
In this subsection, the properties of Dirac plasmons will be examined for all the three materials. Since such plasmon excitations are present only at finite electron/hole dopings, they are also termed as ``extrinsic plasmons". In this work, we consider four different charge-carrier concentrations, including both electron and hole dopings, with $\pm$0.05 and $\pm$0.1 electrons/cell. To achieve these extrinsic conditions, one can shift the Fermi level above or below the Dirac point
until the electron or hole concentrations reach the desired level. A summary of the considered 
doping levels, the corresponding energy shifts and Fermi vectors are given in Table~\ref{table:doping}. 
\begin{table}
\centering
\begin{tabular}{ccccc}
\hline
\hline
doping (electrons/cell) & 0.05  & -0.05 & 0.1 & -0.1  \\
\hline
$E_F$ (graphene) \ &0.0676 \ &-0.0705 \ &0.0916 \ &-0.0982 \\
$k_F$ (graphene) \ &0.129 \ & 0.127 \ &0.175 \ &0.177 \\
$E_F$ (silicene) \ &0.0266 \ &-0.0282 \ &0.0359 \ &-0.0394 \\
$k_F$ (silicene) \ &0.079 \ & 0.078 \ &0.107 \ &0.109 \\
$E_F$ (germanene) \ &0.0252 \ &-0.0263 \ &0.0344 \ &-0.0365\\
$k_F$ (germanene) \ &0.077 \ & 0.077 \ &0.105 \ &0.107 \\
\hline
\end{tabular}
\caption{Corresponding Fermi energies $E_F$ (Ry) and Fermi vectors $k_F$ (/\AA) for graphene, silicene and germanene at the given doping levels.}
\label{table:doping}
\end{table}

\begin{figure}[ht]
\centering
\includegraphics[width=1\textwidth]{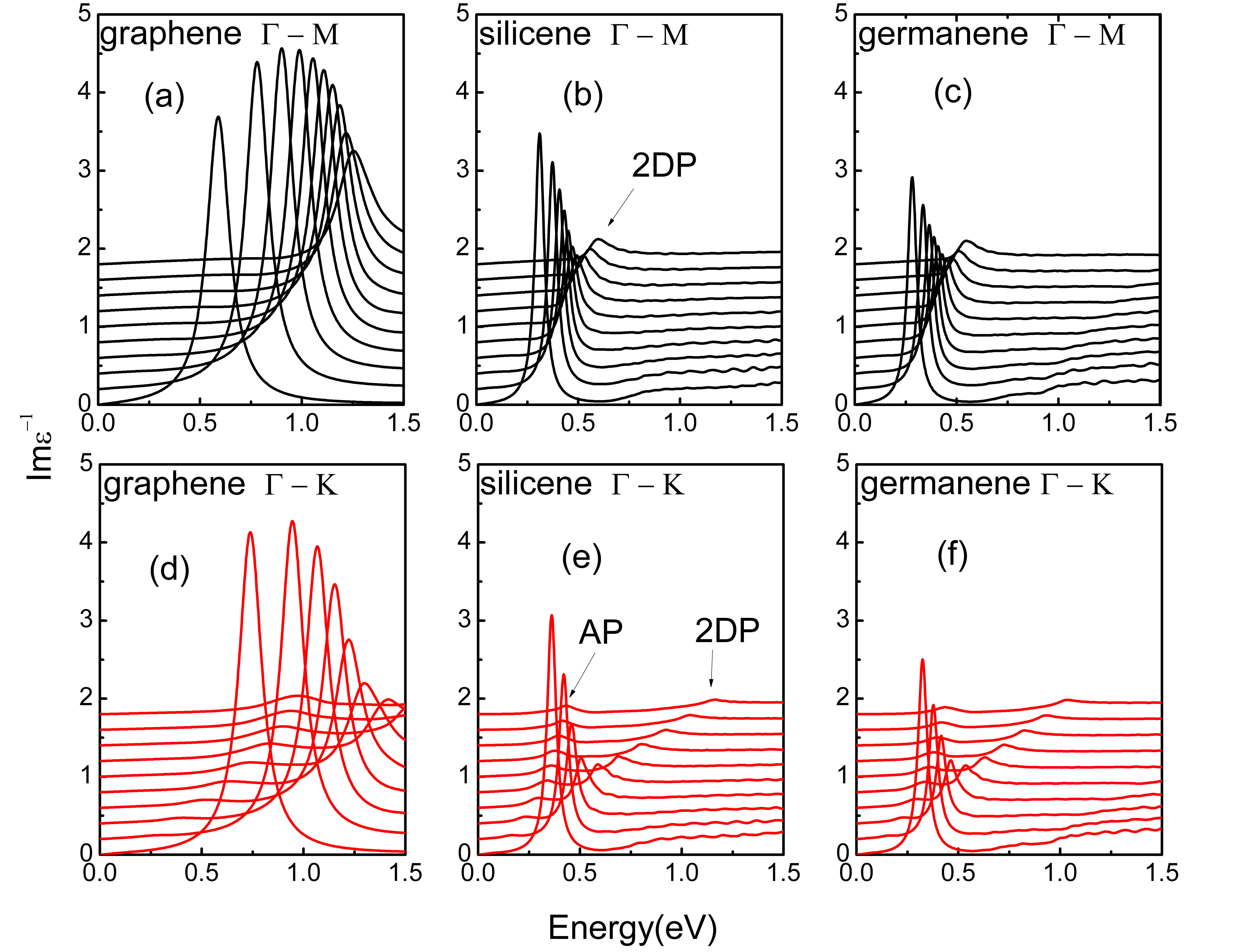}
\caption {The energy-loss spectra of graphene(a)(d), silicene(b)(e) and germanene(c)(f) under 0.05 electrons/cell concentration. The q points lies in the range from 0.014/\AA \ to 0.14/\AA \ and 0.025/\AA \ to 0.25/\AA \ along $\Gamma-M$ and $\Gamma-K$ directions, respectively. One clearly observes that except for a conventional 2D Dirac plasmon (2DP) peak, there exsits another acoustic plasmon (AP) peak in lower energy regime just along $\Gamma-K$ direction.} 
\label{fig:E_loss}
\end{figure}

The energy-loss function $E_{Loss}$ calculated for graphene, silicene and germanene along $\Gamma-M$ and $\Gamma-K$ directions under 0.05 electrons/cell doping are shown in Fig.~\ref{fig:E_loss}. The existence of conventional 2D Dirac plasmons (2DP) along both directions, as well as an additional acoustic plasmon (AP) mode along the $\Gamma$-K direction arising from oscillations of charge carriers with two different Fermi velocities\cite{pines1956electron}, is observed, verifying results reported in previous works for both graphene \cite{pisarra2014acoustic} and silicene \cite{gomez2017plasmon}. 
Here, we demonstrate that similar Dirac plasmon structures also exist in germanene. Furthermore, one can clearly observe that under the same doping level, the intensity of the Dirac plasmons becomes weaker for increasing atomic number.

\subsubsection{Dispersion relationship}
\begin{figure}[ht]
\centering
\includegraphics[width=1\textwidth]{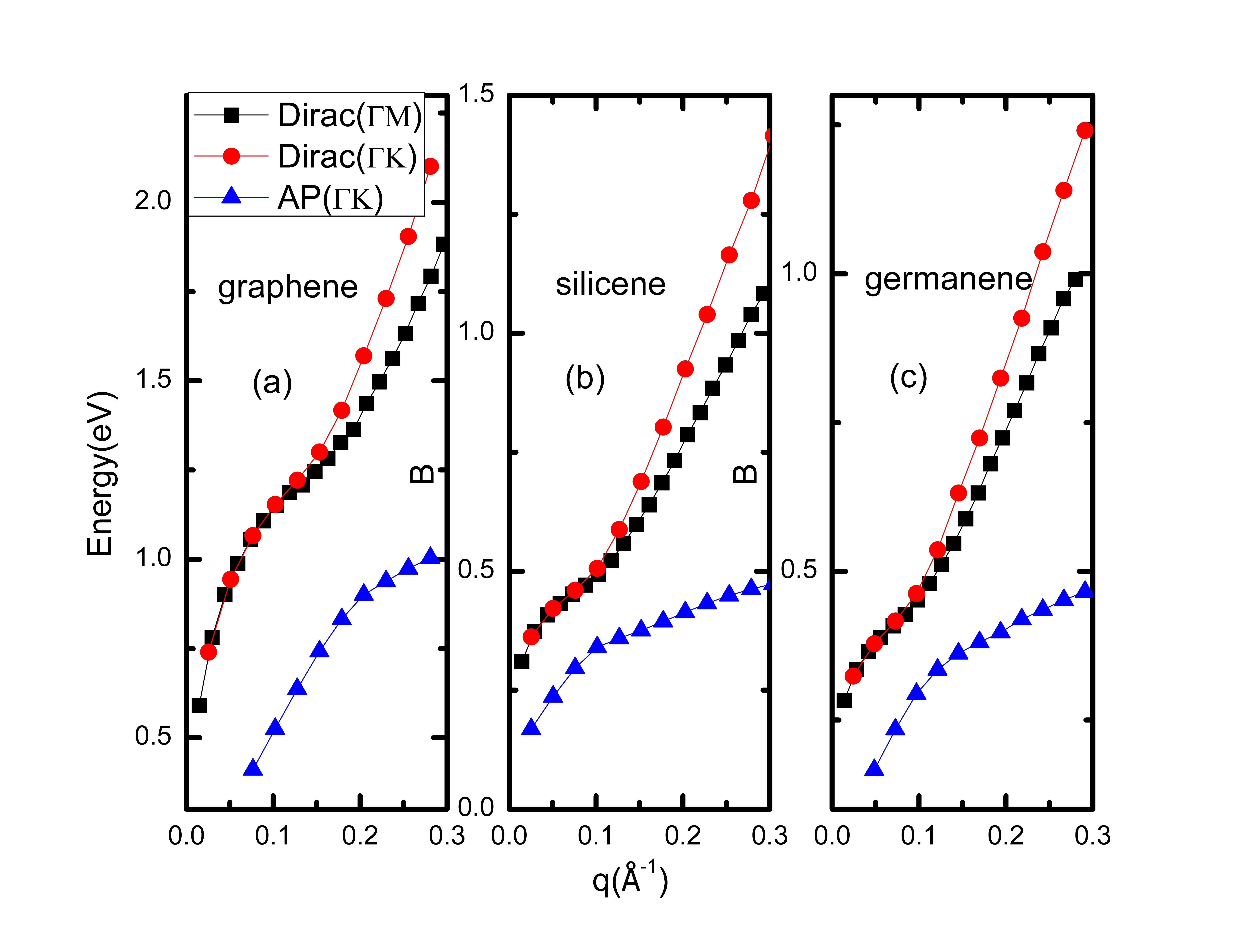}
\caption {The extrinsic plasmon dispersion behavior along $\Gamma-M$ and $\Gamma-K$ directions for graphene(a), silicene(b) and germanene(c) under 0.05 electrons/cell concentration.} 
\label{fig:Gamma_M_Gamma_K}
\end{figure}

\textit{Overall features}
To understand better the dispersion relations of Dirac plasmons, plasmon peak positions as a function of the momentum transfer $q$ are plotted in Fig.~\ref{fig:Gamma_M_Gamma_K} \ along both $\Gamma-M$ and $\Gamma-K$ directions. The behaviors of the Dirac plasmons for the three materials are very similar across the board. It must be remarked that, for conventional 2D Dirac plasmons (2DP) 
outside single-particle excitation (SPE) region, the dispersion roughly follows a $\sqrt{q}$ behavior. Within single-particle excitation region, where the Landau damping comes into play, the conventional Dirac plasmons actually display a quasilinear dispersion behavior, in agreement with what was originally found in model\cite{hwang2007dielectric} and experimental\cite{liu2008} studies. We also notice that the non-coincident 2DP dispersion relations along $\Gamma-M$ and $\Gamma-K$, in particular, the latter also has a unique AP plasmon, which reflects the anisotropicity of plasmons along different directions.

\begin{figure}[ht]
\centering
\includegraphics[width=1\textwidth]{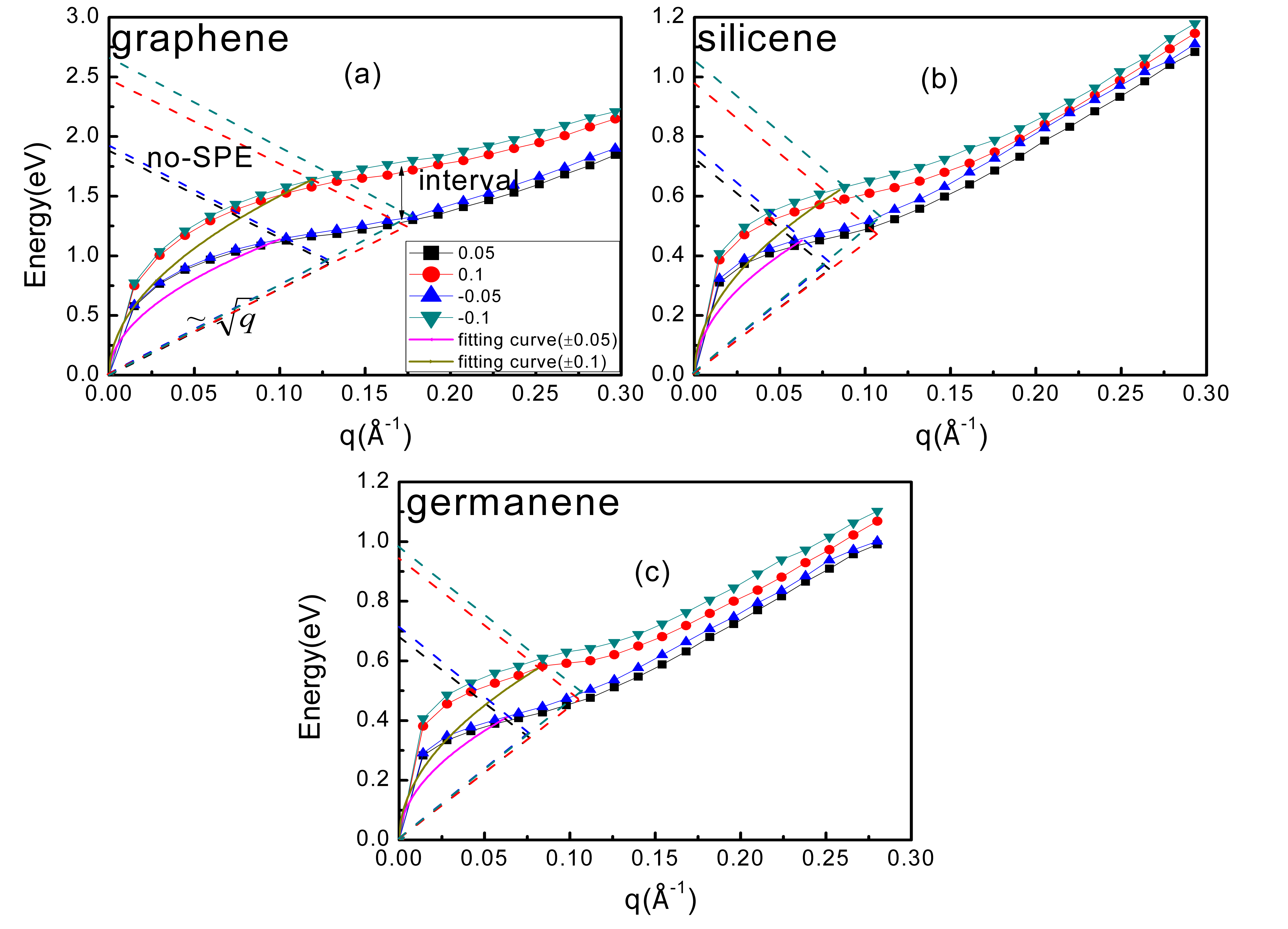}
\caption {The 2D Dirac plasmon dispersions of graphene(a), silicene(b) and germanene(c) under $\pm$0.05 and $\pm$0.1 dopings along $\Gamma-M$ direction. The black, blue, red and green dashed lines represent the single-particle excitation(SPE) boundaries under 0.05, -0.05, 0.1 and -0.1 electrons/cell doping concentrations, respectively. The pink and brown lines within no-SPE region correspond to $\sqrt{q}$ dispersion of free 2D electron gas. One can clearly observe the $\sqrt{q}$ hehavior without SPE and quasilinear behavior within SPE.} 
\label{fig:doping_dispersion}
\end{figure}

\textit{Dispersion under different doping levels.}
In order to find out the influence when different doping concentrations involved in, the dispersion relationship of 2DP under four charge-carrier concentrations, $\pm$0.05 and $\pm$0.1 electrons/cell, for graphene, silicene and germanene are depicted in Fig.~\ref{fig:doping_dispersion}. For convenience, we only cover the results along $\Gamma-M$ direction. In particular, single-particle excitation boundaries are shown in Fig.~\ref{fig:doping_dispersion}. One can define a cutoff vector $q_c$, determined by the intersection point of the dispersion curve and the boundary of SPE region, indicates the position where plasmon enters the damping region, namely, when $q<q_c$, the plasmon dispersion locates in no-SPE region, while $q>q_c$, it lies in the SPE region. The calculated $q_c$ under different dopings and materials are shown in Table~\ref{table:qc}. 
\begin{table}
\centering
\begin{tabular}{ccccc}
\hline
\hline
doping (electrons/cell) & 0.05  & -0.05 & 0.1 & -0.1  \\
\hline
graphene \ &0.104 \ &0.105 \ &0.127 \ &0.131 \\
silicene \ &0.064 \ &0.065 \ &0.083 \ &0.087 \\
germanene \ &0.064 \ &0.064 \ &0.081 \ &0.082\\
\hline
\end{tabular}
\caption{The cutoff vector $q_c$(/\AA) under different dopings and materials.}
\label{table:qc}
\end{table}
It should be stressed that, the fitting curve within the no-SPE region indicates a $\sqrt{q}$ dispersion behavior and almost quasilinear outside this region as observed, no matter for which materials or dopings. Further, by inspection of Fig.~\ref{fig:doping_dispersion}, we witness that with the increase of q, the interval between plasmon dispersion curves at $\pm$0.05 and $\pm$0.1 concentrations increases first, and then decreases, no matter in which kind of materials. The rising process is easy to understand while the descending process originates from the larger $q_c$ in higher doping concentration plasmon dispersion relation, makes it later into to the damping zone where the slope of the curve is steeper, therefore the interval will be shortened. The $q_c$ data can be obtained from Table~\ref{table:qc}. By the way, We also found that the same quantity of concentrations give highly consistent results, it has nothing to do with the doping type(electron/hole doping).

\begin{figure}[ht]
\centering
\includegraphics[width=1\textwidth]{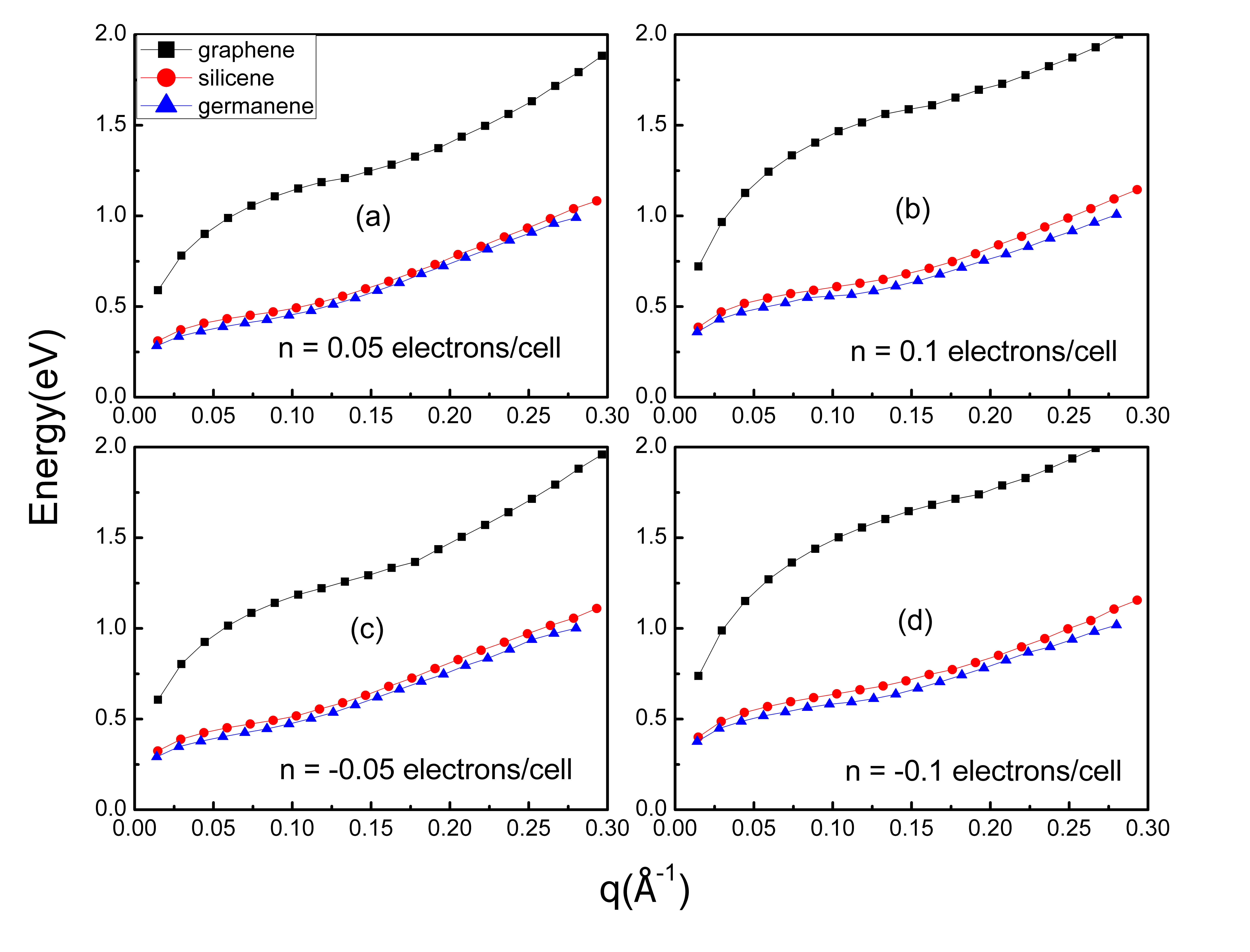}
\caption {The 2D Dirac plasmon dispersions of graphene, silicene and germanene under 0.05(a), 0.1(b), -0.05(c) and -0.1(d) electrons/cell doping concentrations.}
\label{fig:material_dispersion}
\end{figure}

\textit{Dispersion under the same doping but different materials.}
In Fig.~\ref{fig:material_dispersion}, we compare the conventional 2D Dirac plasmon dispersion relation of these three materials - graphene, silicene and germanene, under the same charge-carrier concentrations.  Here, we only cover the results along the $\Gamma-M$ direction for convenience. It should be noted that the dispersion relations will be redshifted as the atomic number increases, no matter in which concentrations. This phenomenon can be attribute to the behavior of ``Dirac cone'' like band structure around K points near $E_F$. The bands in this regime become ``flatter'' for increasing atomic numbers, leading to slighter $E_F$ shifts which can be seen in Table.~\ref{table:doping}, and then cause lower excitation energies. Moreover, it is easy to see, compared to graphene, the dispersion behaviors of silicene and germanene enter the quasilinear region earlier, and this outcome originates from their smaller cutoff vector $q_c$.

\subsubsection{Lifetimes}
The lifetime of plasmons is an important quantity to be considered for technological applications. In this work, our first-principles lifetime calculations are restricted to the electron-electron scattering at the RPA level, and the effects from phonons, impurities, and disorders\cite{jablan2009plasmonics,mermin1970lindhard,principi2014plasmon}, 
as well as electronic effects beyond RPA\cite{principi2013intrinsic}, are not considered. Theoretically, the lifetime of plasmon excitations is inversely proportional to the width of the spectral peaks, and hence the calculation of lifetime amounts to determining the linewidth of the plasmon spectral peaks. The full widths at half maximum (FWHM) of the plasmon peaks is used to determine the lifetimes. By inspection of Fig.~\ref{fig:E_loss}, it is noted that as $q$ increases, these Dirac plasmon peaks quickly get broadened with the magnitude diminishes. But at the same time, we found that the linewidth of the Dirac plasmon peaks depends on the broading parameter $\eta$ which is used in calculations. To put it more strictly, our method to get the accurate FWHM is given in Ref.~\cite{li2017first}. Compared to previous work, which is exhibited in Appendix.~\ref{sec:eta}, not only the effect on plasmon lifetimes, but also the plasmon excitation energies (peak positions) under different values of $\eta$ are involved in.

\begin{figure}[ht]
\centering
\includegraphics[width=1\textwidth]{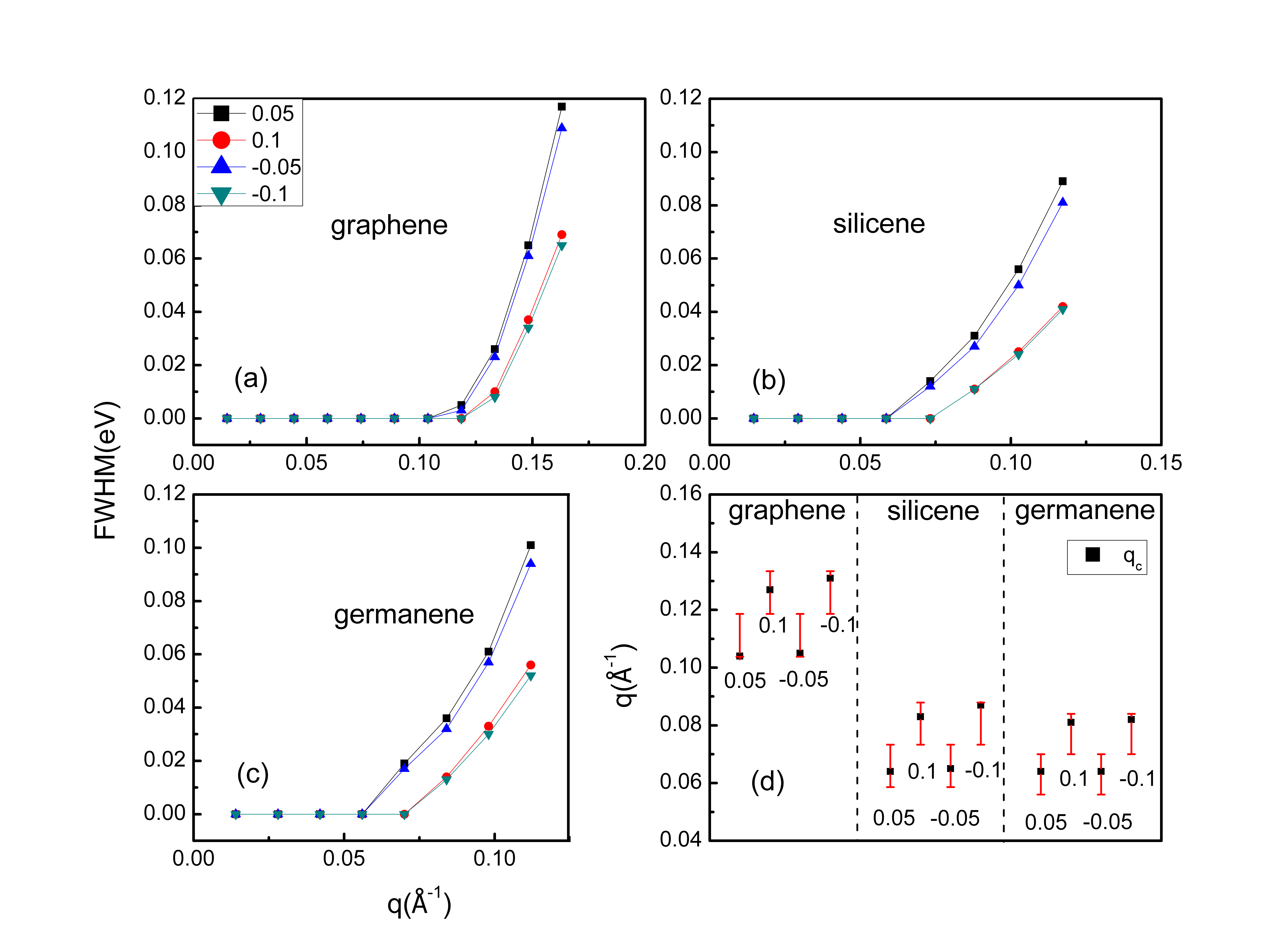}
\caption {The FWHM of 2D Dirac plasmon peaks for graphene(a), silicene(b) and germanene(c) under $\pm$0.05 and $\pm$0.1 doping concentrations. (d) the range of the lifetime transition point(red bar) compared to cutoff vectors $q_c$(black dot) derived from Table~\ref{table:qc}. From left to right shows the result for graphene, silicene and germanene, respectively, and in each part, the datas are arranged in 0.05, 0.1, -0.05 and -0.1 electrons/cell order.} 
\label{fig:lifetime1}
\end{figure}

\textit{Lifetimes under different dopings.}
In this part, our intention is to achieve the influence of different doping concentration on plasmon lifetimes. The FWHMs as a function of momentum transfer q along $\Gamma-M$ direction are depicted in Fig.~\ref{fig:lifetime1} for graphene(a), silicene(b) and germanene(c), respectively. Generally, in the region with small q values, the FWHM is strictly zero and becomes larger for increasing q outside, namely, the lifetime is infinite within the zone and gets finite without the zone. Despite the limitation by the calculation conditions, one can not get the accurate critical transition point, but we have reasons to believe this transition point is exactly the cutoff vector $q_c$ as defined above. As is shown in Fig.~\ref{fig:lifetime1}(d), in all cases, the cutoff vectors $q_c$ obtained from Table~\ref{table:qc} locate within the desirable range, from the last zero q-point to the first non-zero q point which can be extracted from Fig.~\ref{fig:lifetime1}(a)(b)(c). It should be stressed that, the lifetime at the RPA level is infinite in no-SPE and finite in SPE region, and it becomes shorter as q increases, which is consistent with previous work\cite{wunsch2006dynamical}. Moreover, one clearly observes that the FWHMs get smaller for increasing doping concentrations at the same q for all three materials, meaning the lifetime becomes longer as the doping concentration increases. This phenomenon can be attribute to their larger values of $q_c$, which make them later into the damping zone.

\begin{figure}[ht]
\centering
\includegraphics[width=1\textwidth]{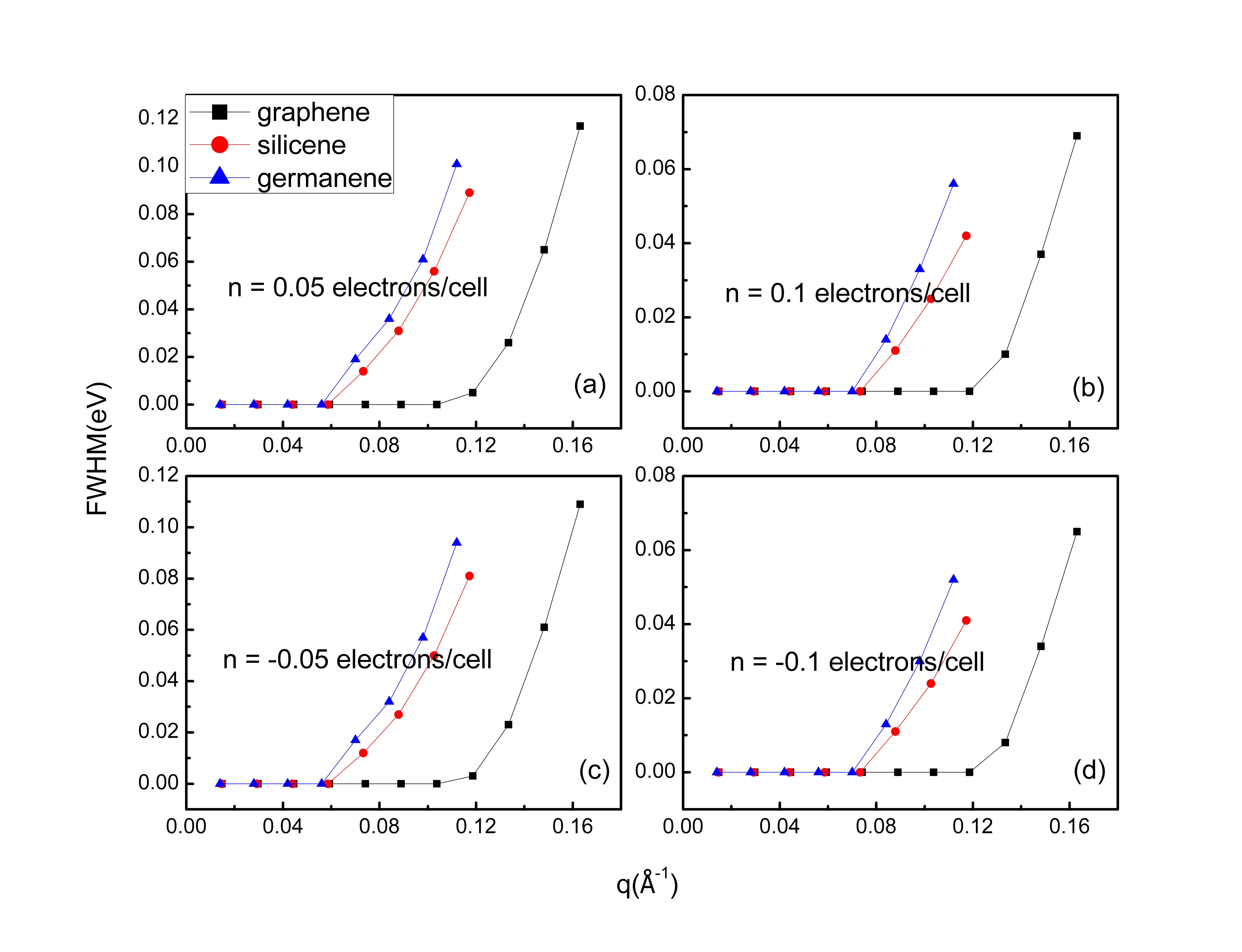}
\caption {The FWHM of 2D Dirac plasmon peaks for graphene, silicene and germanene under 0.05(a), 0.1(b), -0.05(c) and -0.1(d) doping concentrations.} 
\label{fig:lifetime2}
\end{figure}

\textit{Lifetimes for different materials under the same doping.}
In Fig.~\ref{fig:lifetime2}, we compare the conventional 2D Dirac plasmon FWHMs of the three materials - graphene, silicene and germanene, under the 0.05(a), 0.1(b), -0.05(c) and -0.1(d) electrons/cell doping concentrations along $\Gamma-M$ direction. It is easy to see that no matter in which concentration, the FWHM becomes larger for increasing atomic numbers, meaning shorter lifetimes. This result can also be explained by the values of $q_c$. As atomic number increases, the cutoff vector $q_c$ decreases and then lead to a shorter lifetime.

\subsection{Intrinsic plasmons}
\begin{figure}[ht]
\centering
\includegraphics[width=1\textwidth]{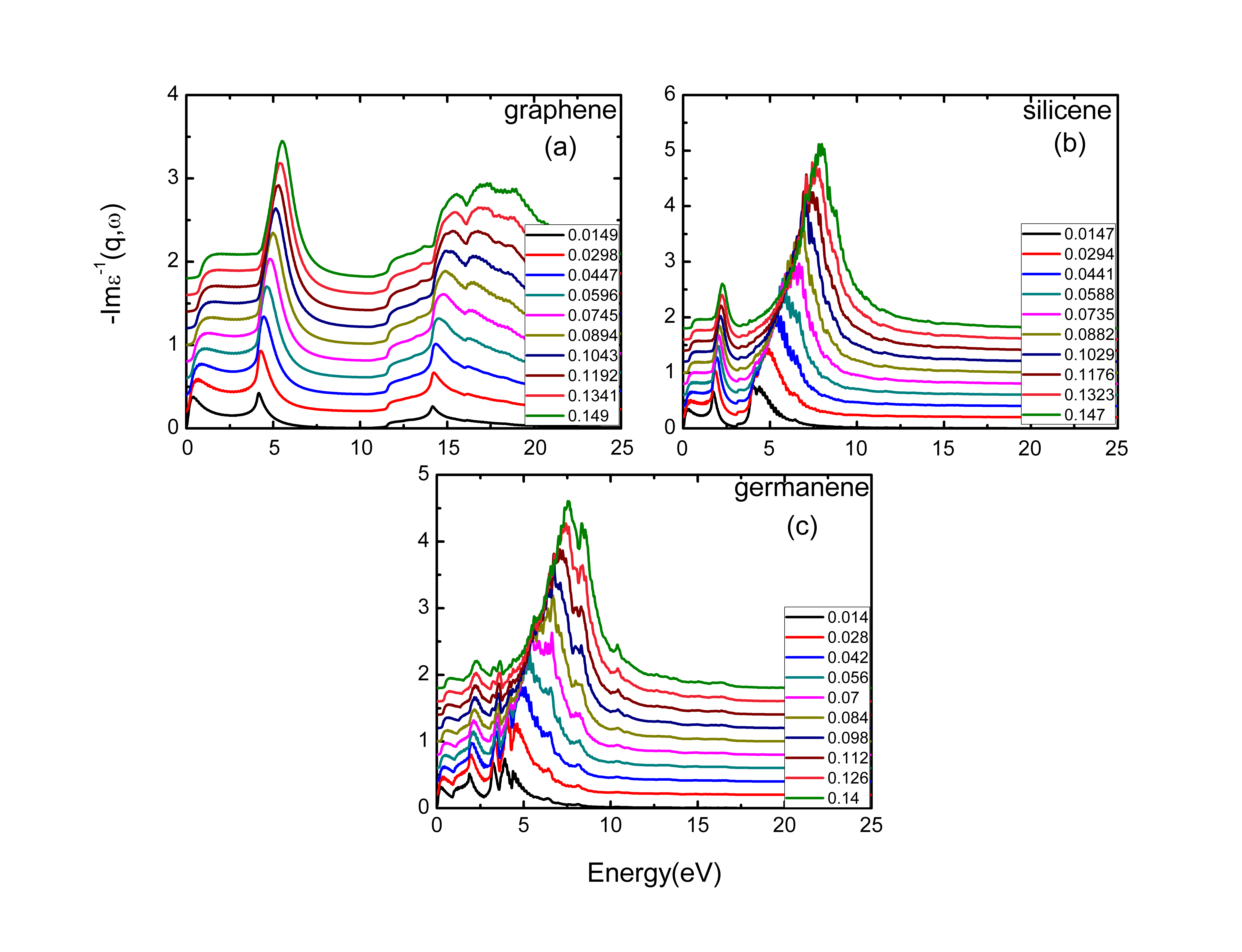}
\caption {The overall electron energy loss spectroscopy (EELS) of graphene(a), silicene(b) and germanene(c) along $\Gamma-M$ direction. The q points lies in the range from 0.014/\AA \ to 0.14/\AA.} 
\label{fig:intrinsic_overall}
\end{figure}

The energy-loss spectras of undoped (intrinsic) graphene, silicene and germanene along $\Gamma-M$ direction are  depicted in Fig.~\ref{fig:intrinsic_overall}. By inspection of Fig.~\ref{fig:intrinsic_overall}, in broad terms, two main plasmon structures can be clearly distinguished, i.e., a $\pi$-like plasmon branch and a $\pi$-$\sigma$ plasmon branch. Within the $q < 0.15$/\AA \ regime, the $\pi$-like plasmons are located at 4.0-5.5 eV, 1.7-2.3 eV, 1.8-2.3 eV, and the $\pi$-$\sigma$ plasmons are located at 14-20 eV, 4-8 eV, 4-8 eV, for graphene, silicene, and germanene, respectively. Compared to graphene, it is noted that the plasmon peaks of silicene and germanene are redshifted, which can be attributed to the shrinking band gap between $\pi$ to $\pi^\ast$ or $\sigma^\ast$ bands at M point. Besides, we also observed that in graphene, the $\pi$ and $\pi$-$\sigma$ plasmons have similar peak intensities \cite{eberlein2008plasmon}, while the $\pi$-$\sigma$ peak of silicene is generally larger than the $\pi$-like peak, which is even more obvious in germanene. This phenomenon can be ascribed to the weakening of the $\pi$ bands in silicene \cite{gomez2017plasmon} and germanene due to the mixed $sp^2$-$sp^3$ hybridization.

\begin{figure}[ht]
\centering
\includegraphics[width=1\textwidth]{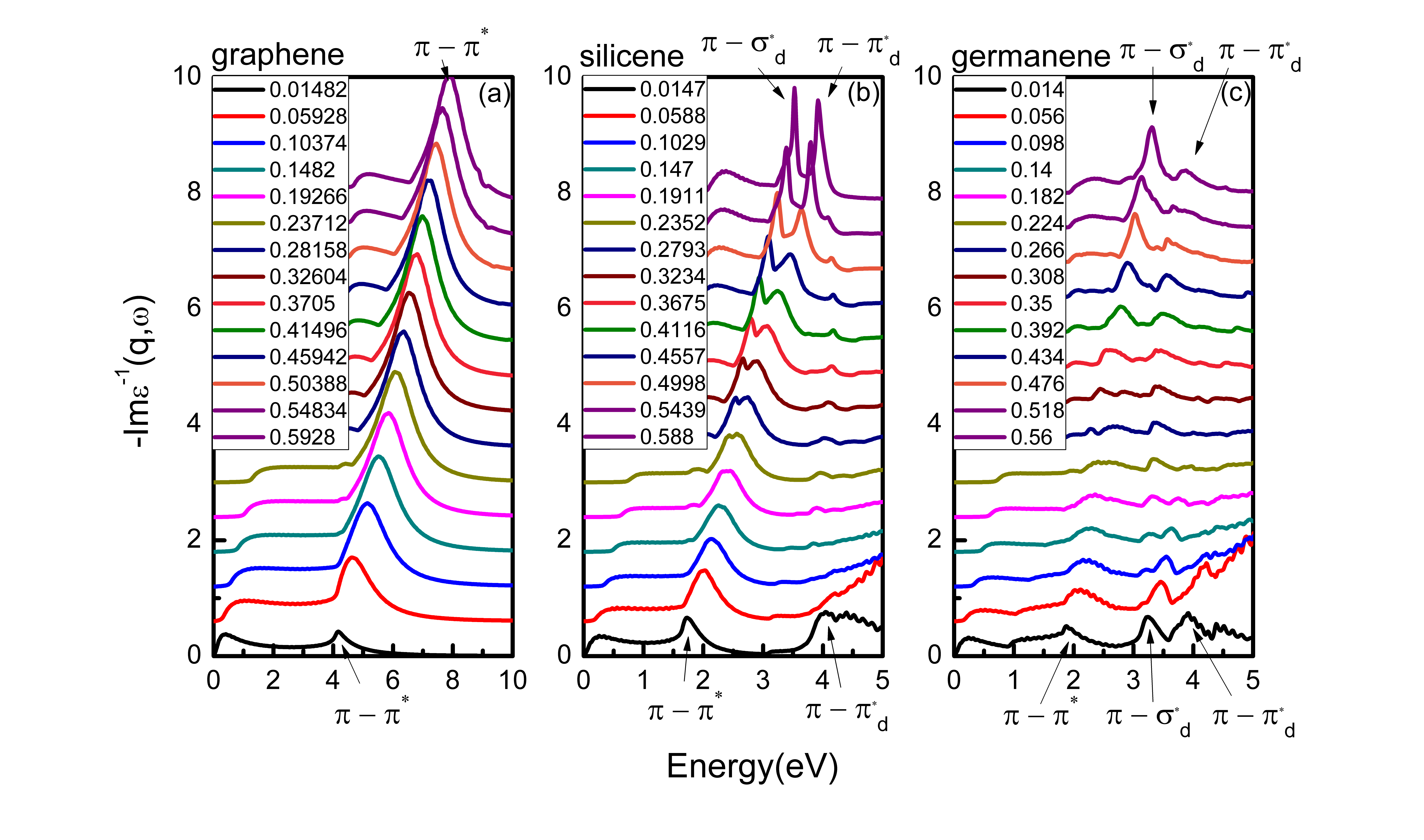}
\caption {The electron energy loss spectroscopy (EELS) of intrinsic graphene(a), silicene(b) and germanene(c) in smaller energy window along $\Gamma-M$ direction. The corresponding transitions for peaks are labeled thereon.} 
\label{fig:intrinsic_dispersion}
\end{figure}

In this work, we mainly concern about the $\pi$-like plasmon in lower energy regime. In order to get a better view of the details, the energy loss functions of graphene, silicene and germanene in smaller energy window along the $\Gamma-M$ direction are further presented in Fig.~\ref{fig:intrinsic_dispersion}. One clearly observes that in graphene, $\pi$ plasmons form a single peak structure all the time, while in silicene and germanene, especially at large q values, the one-peak structure splits into a two-peak structure\cite{gomez2017plasmon}. For silicene, in the energy-momentum region $\omega > 2.5$ eV and $q > 0.23$ \AA$^{-1}$, this phenomenon starts to appear, while for germanene, it exsits all the time. The reason can be attributed to the complex and hybridized band structure of silicene and germanene, as is shown in Fig.~\ref{fig:band_dos}. In the energy window 0-5 eV, for graphene, few $\sigma^\ast$ bands come into play, thus leading to a pure $\pi$ $\rightarrow$ $\pi^\ast$ peak. While for silicene and germanene, $\sigma^\ast$ bands are involved in and making outstanding contributions to form the $\sigma^\ast_d$ and $\pi^\ast_d$ bands which can be clearly seen from the PDOS peaks in Fig.~\ref{fig:band_dos}(e)(f). The transitions $\pi$ $\rightarrow$ $\sigma^\ast_d$ and $\pi$ $\rightarrow$ $\pi^\ast_d$ form the so-called two-peak structure. Besides, the reason why we can not see this spilt 
in silicene at small $q$'s is caused by the fact that large $\pi^\ast_d$ peak hides the $\sigma^\ast_d$ contribution, which can be seen in Fig.~\ref{fig:band_dos}(e) TDOS(black lines), barely visible $\sigma^\ast_d$ peak. However, this situation does not occur in germanene.

\section{\label{sec:conclusion}Summary}
In this work, first, some discussions are made on computational details to make sure the correctness of calculations, including the formula to construct the inverse of the dielectric function, with the 
Coulomb kernel selects the 2D-form, not the gradual-form, and the demonstration of the rationality for using the peak positions of Im($\epsilon^{-1}$) to describe the plasmon excitation energies. Then, we performed a systematic study of the Dirac plasmons present in graphene, silicene, and germanene under finite dopings and found that, as the atomic number gets larger, the plasmon excitation energies will get lower, with weakening plasmon intensities and decreased plasmon lifetimes. Furthermore, for all three materials, the lifetimes of the Dirac plasmons
will become longer as the doping concentration increases.  Most importantly, through data analysis, we demonstrate that, in the region without single-particle excitation (SPE), the plasmon dispersion shows a $\sqrt{q}$ behavior and the lifetime is infinite at the RPA level, while in the single-particle excitation region, the plasmon dispersion shows a quasilinear behavior and the lifetime is finite. Moreover, we also examined the behavior of the intrinsic plasmons for these three materials in their pristine forms, and found that the $\pi$-like plasmons are redshifted and become weaker as the atomic number increases. Unlike what happens in graphene, the energy loss functions of silicene and germanene show a two-peak structure. In particular, the double-peak structure emerges for germanene even at small $q$'s. This phenomenon can be traced back to the hybridization of the $\pi$ and $\sigma$ electrons, arising from the buckled structure of silicene and germanene.

\begin{appendix}
\section{\label{sec:discribe_2D}Which formula is better to discribe the 2D dielectric function?}
In Eq.~(\ref{eq:inverse_epsilon}), there are two different choices to determine the 2D dielectric function, the pure 2D-form and the gradual-form, which can be used to calculate the inverse of the dielectric function as, 
\begin{equation}
   \epsilon^{-1} = 1 + \frac {2\pi L_z} {\bar{q}} \chi 
  \label{eq:first}
\end{equation}

or 
\begin{equation}
	\epsilon^{-1} = 1 + \frac {4\pi(1-e^{-\bar{q}L_z/2})} {{\bar{q}}^2} \chi 
  \label{eq:second}
\end{equation}

\begin{figure}[ht]
\centering
\includegraphics[width=1\textwidth]{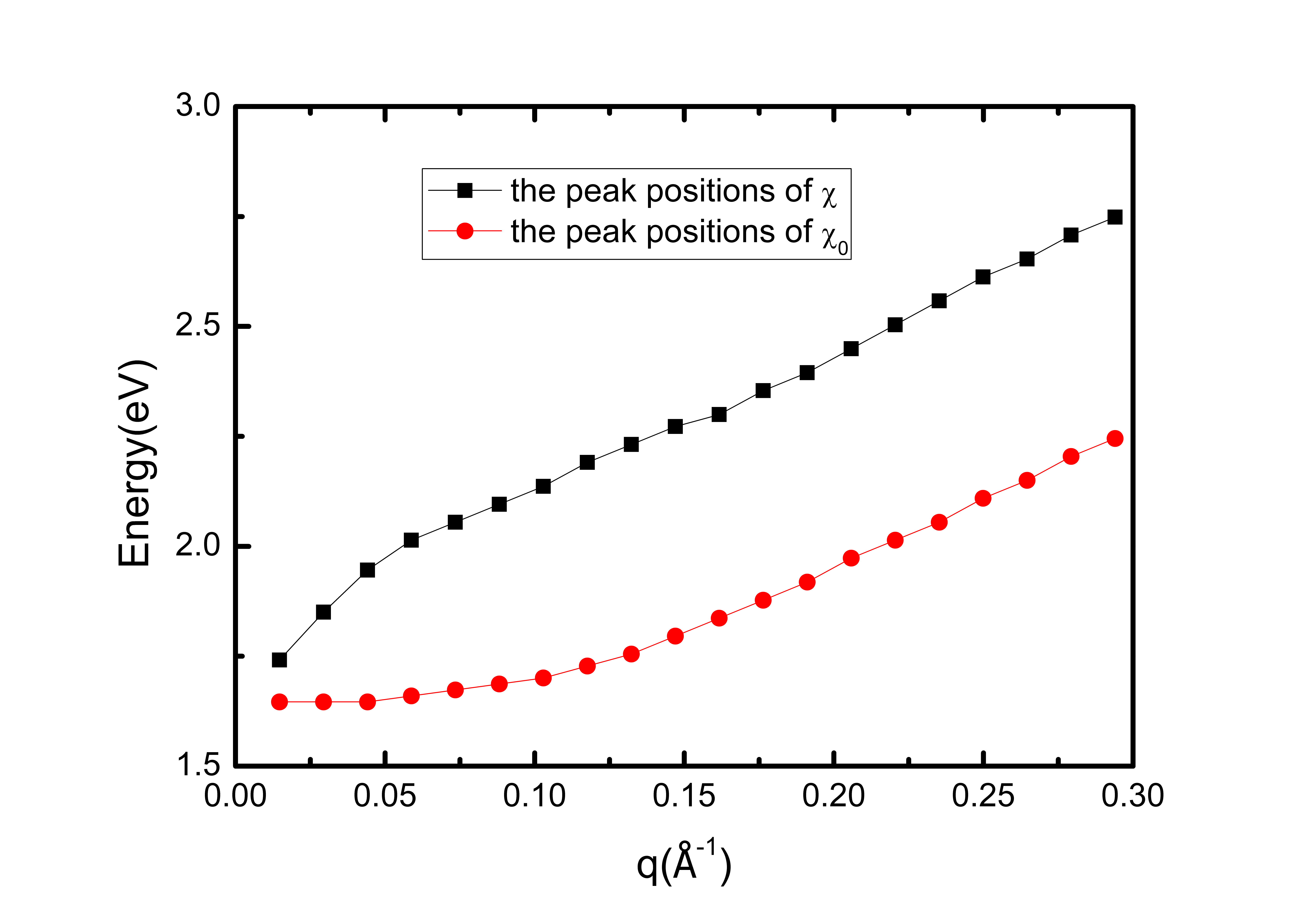}
\caption {Dispersion behavior of the peak positions as a function of q, extracted from the $\pi$-like peaks in Im$\chi_{00}$ and Im$\chi^0_{00}$ for silicene along $\Gamma-M$ direction.}
\label{fig:chi0_chi}
\end{figure}

But which one is the better choice for the 2D dielectric function discription? To solve this problem, we'd better return to its physical nature. For these IV-group element based hexagonal honeycomblike structures, whether the $\pi$ peaks are collective plasmon excitations or single-particle interband transitions has given rise to much controversy\cite{nazarov2015electronic,nelson2014electronic}. The most efficient method to  distinguish if these peaks are plasmons or not is by examining the pole structures of the interacting response function $\chi$ and the non-interacting response function $\chi_0$. 
In Fig.~\ref{fig:chi0_chi}, we compare the dispersion relationships of the peaks in both Im$\chi_0$ and Im$\chi$ for silicene along $\Gamma-M$ direction. Peak positions in Im$\chi_0$, derived directly from the band structure, reflects single-particle interband transitions, whereas those in Im$\chi$, affected by the long-range Coulomb interactions between different particle-hole pairs, could reflect the collective behavior. One clearly observes that, as q increases, the dispersion relations of collective excitation and individual transition energies are radically different. While peak energies in Im$\chi_0$ show a very slow increase as q increases (almost dispersionless for small q’s), those in Im$\chi$ display a $\sqrt{E^2_g + \beta q}$ behavior, as proposed in Ref.~\onlinecite{li2017first}. Graphene and germanene give the similar results. This is a direct demonstration that the $\pi$ peaks are real plasmon excitations. 

\begin{figure}[ht]
\centering
\includegraphics[width=1\textwidth]{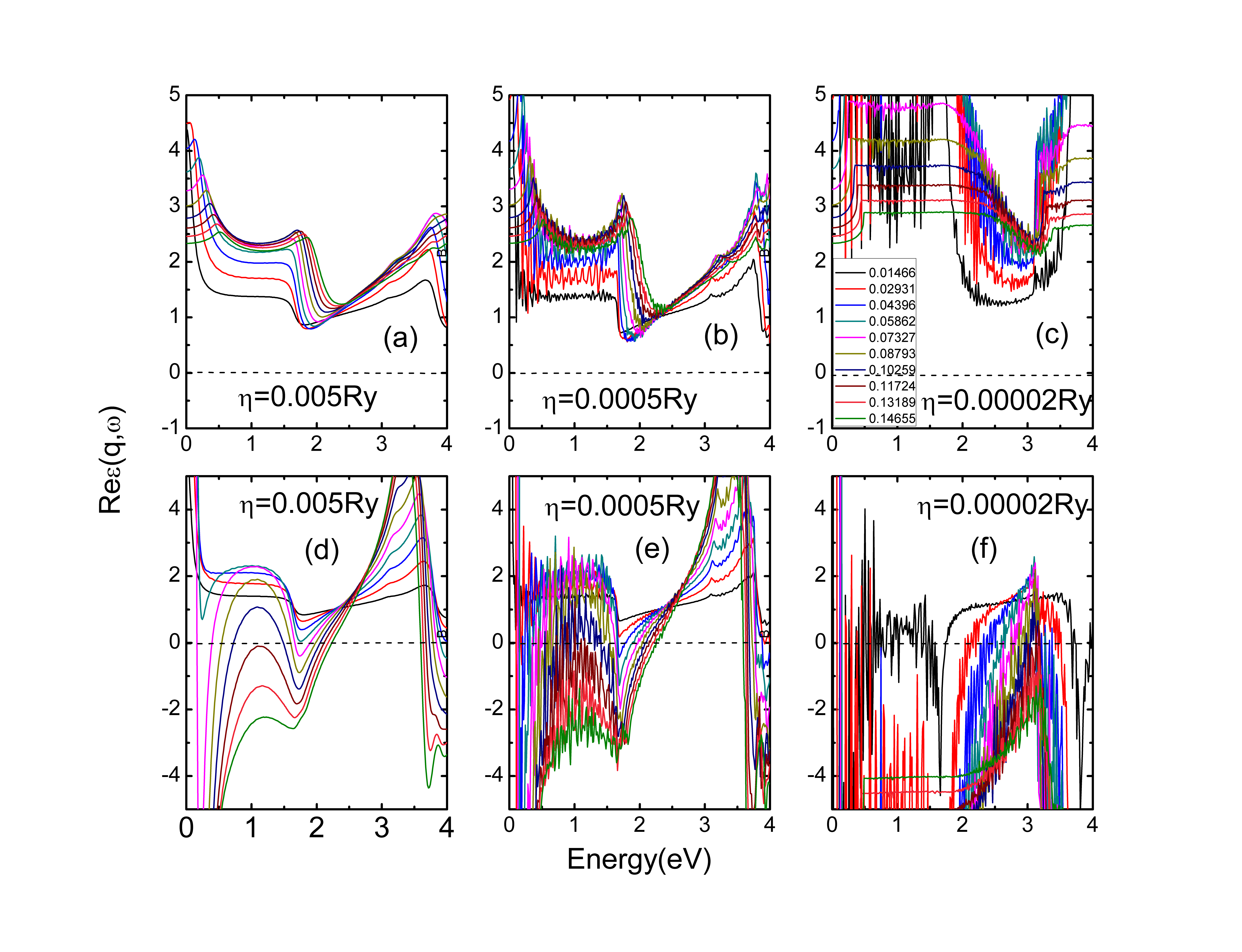}
\caption {the real part of the dielectric function (Re$\epsilon$) for silicene along $\Gamma-M$ direction. From left to right, $\eta$ is set to be 0.005Ry, 0.0005Ry and 0.00002Ry, respectively. The upper panels(a)(b)(c) shows the result of using gradual-form(Eq.~(\ref{eq:second})) and the lower panels(d)(e)(f) shows the result of using 2D-form(Eq.~(\ref{eq:first})). The q points along $\Gamma-M$ direction lies in the range from 0.014/\AA \ to 0.14/\AA.} 
\label{fig:epsilon_method}
\end{figure}

Once we've determined that these $\pi$ peaks are real plasmon excitations, then begin the next step. A clear criterion for an excitation to be classified as plasmon is the requirement of real-part of the dielectric function to cross zero at the corresponding energy. The real-part of the dielectric function of silicene with different broadening parameter $\eta$ obtained by both methods are  depicted in Fig.~\ref{fig:epsilon_method}. One clearly observes that, if the gradual-form is used, even as we artificially approach $\eta$ $\rightarrow$ 0, it can not reach Re($\epsilon$) = 0 conditions. Meanwhile, we also use another gradual-form $v$ as a comparation\cite{gomez2017tunable}, which is formed as
\begin{equation}
v_{\bf{G},\bf{G}^\prime} = \frac {2\pi} {|\bf{q} + \bf{g}|} \int_{-L/2}^{L/2}dz \int_{-L/2}^{L/2}dz^\prime e^{i(G_zz-G^\prime_zz^\prime)-|\bf{q}+\bf{g}||z+z^\prime|}
\label{eq:2D_coulomb_kernel_3}
\end{equation}
and gives similar results. On the contrary, if 2D-form is used, despite it can not satisfy the Re($\epsilon$) = 0 condition at small q values when $\eta$ is large, this situation will soon change when smaller $\eta$ is used. From what has been discussed above, Eq.~(\ref{eq:first}) may be the better choice. Furthermore, as all the datas come from the $E_{Loss}$, in other words, the imaginary part of the $\epsilon^{-1}$, both methods gives similar results, by just a multiple relationship, namely, both schemes give the equal plasmon excitation energies and lifetimes.

\section{\label{sec:eta}The influence of different values of $\eta$ on calculations}
A positive broadening parameter $\eta$ is used in Eq.~(\ref{eq:chi_0}) to avoid singularity. Theoretically, ignore the amount of computation, one can get the accurate results on $\eta = 0$ with infinite k points. In this part, we will discuss the influence on the calculation results when using different non-zero $\eta$, mainly focus on the plasmon excitation energies(dispersion peak positions) and lifetimes(FWHM of the dispersion peak).

\begin{figure}[ht]
\centering
\includegraphics[width=1\textwidth]{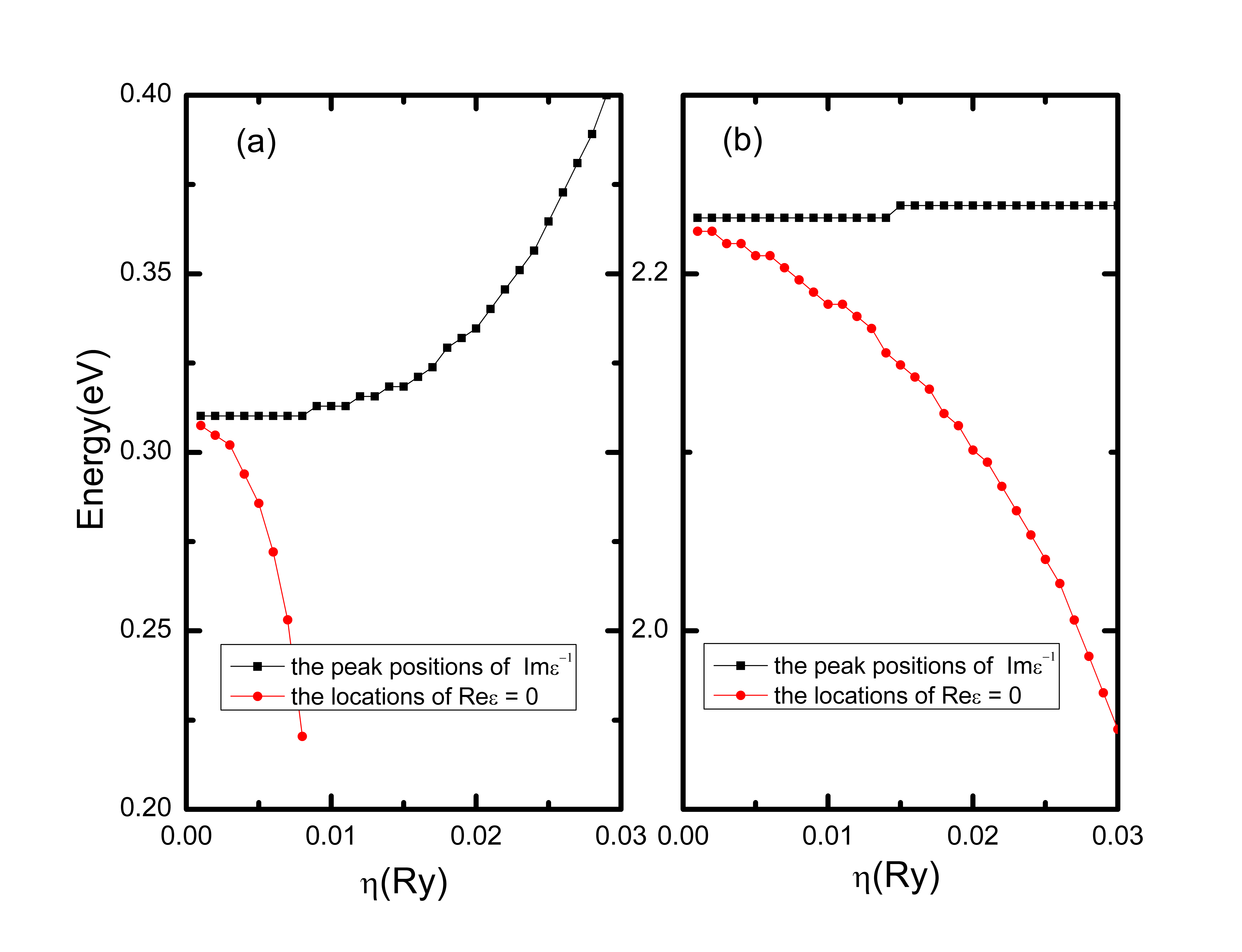}
\caption {the peak positions of Im($\epsilon^{-1}$) and the Re($\epsilon$) = 0 locations with different $\eta$ values for silicene 2D Dirac plasmon at q = 0.0147/\AA \ (a), and for silicene $\pi$-like plasmon at q = 0.1341/\AA \ (b) along $\Gamma-M$ direction.}
\label{fig:eta_convergence}
\end{figure}

First, we consider its influence on the plasmon excitation energies. As we all know, the plasmon excitation energy $\omega$ is determined under Re($\epsilon$($q$,$\omega$)) = 0 conditions. Sometimes, the peak positions of Im($\epsilon^{-1}$($q$,$\omega$)) can also be treated as a criterion. In Fig.~\ref{fig:eta_convergence}, we compare the peak positions of Im($\epsilon^{-1}$($q$,$\omega$)) as well as the Re($\epsilon$($q$,$\omega$)) = 0 locations at a given $q$ point when artificially approach the broadening parameter $\eta$ to zero both for 2D Dirac and $\pi$-like plasmon of silicene. It is noted that, no matter for which kind of plasmon, the peak positions of Im($\epsilon^{-1}$) and the Re($\epsilon$) = 0 locations tend to be consistent when $\eta$ $\rightarrow$ 0, which represents the true plasmon excitation energy. Compared to the Re($\epsilon$) = 0 locations, the peak positions of Im($\epsilon^{-1}$) are easier to converge. As a consequence, we use the peak positions of Im($\epsilon^{-1}$) to determine the corresponding excitation energies. 

\begin{figure}[ht]
\centering
\includegraphics[width=1\textwidth]{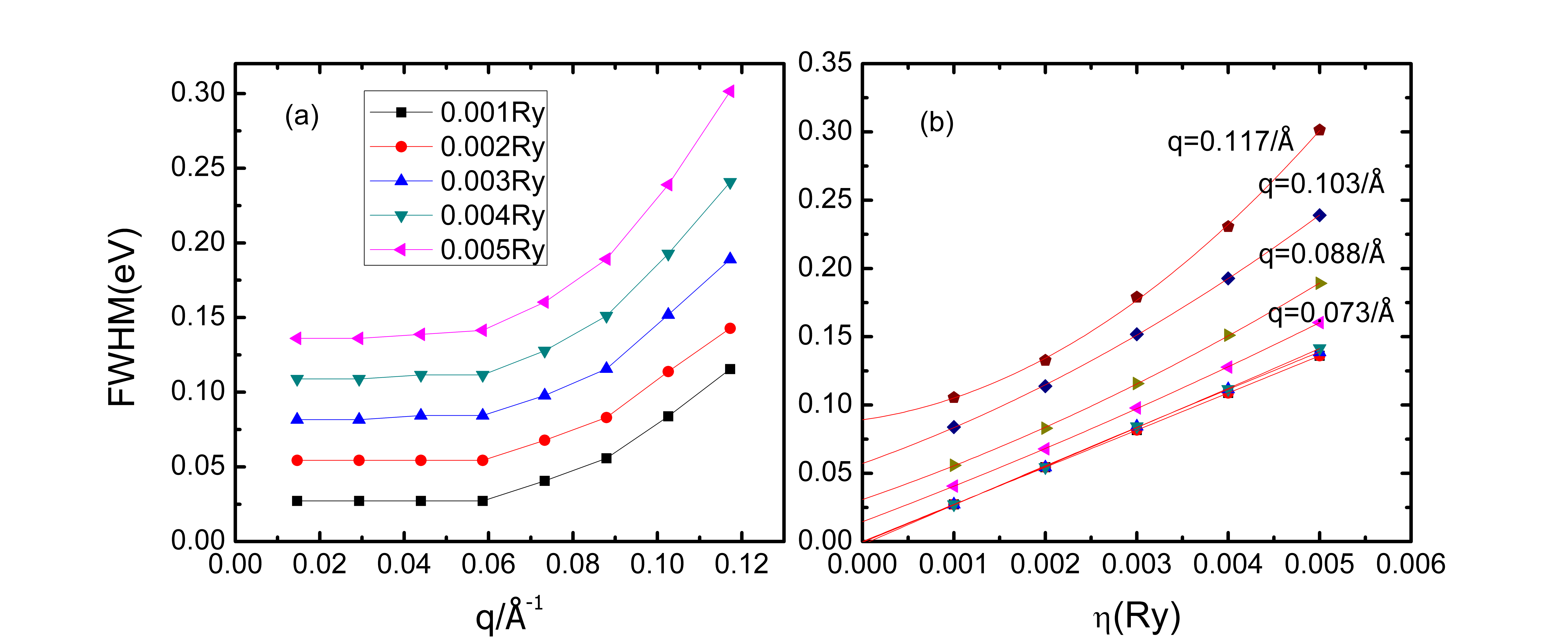}
\caption {(a) The FWHM of the conventional 2D Dirac plasmon peaks as a function of q along the $\Gamma-M$ direction for different $\eta$ values from 0.001Ry to 0.005 Ry. (b) The extrapolating result of FWHM values to the limit $\eta$ $\rightarrow$ 0 for silicene dirac plasmon at 0.05 electrons/cell doping concentration along $\Gamma-M$ direction.}
\label{fig:FWHM}
\end{figure}

Furthermore, we take its influence on plasmon lifetimes into account. It is found that the FWHM of peaks on Im($\epsilon^{-1}$) changes considerably if different $\eta$ values are used. Theoretically, the true FWHM can be obtained if $\eta$ is strictly zero, the steps are as below: (1) calculate the loss spectra for several different $\eta$ parameters and extract the FWHM of each peak. Fig.~\ref{fig:FWHM}(a) shows the FWHM of silicene 2D Dirac plasmon peaks at different q's while $\eta$ is set from 0.001Ry to 0.005Ry. One can clearly see that the FWHM values depend appreciably on the parameter $\eta$. As $\eta$ is reduced, the FWHM value decreases. (2) obtain the final results by extrapolating the FWHM values from finite $\eta$ to $\eta$ = 0. As it shown in Fig.~\ref{fig:FWHM}(b), in the no-SPE regime, the FWHM naturally goes to zero.
\end{appendix}

\section*{Acknowledgments}
P. L. acknowledges the support from the National Natural Science Foundation of China (No.11904284), the Natural Science Basic Research Program of Shaanxi (Program No.2020JQ614) and The Youth Innovation Team of Shaanxi Universities. X. R. acknowledges the support from  the National Natural Science Foundation of China (Grant No. 11874335).
%
\end{document}